\documentclass[showpacs,prl,twocolumn,floatfix,reprint]{revtex4-2}
\usepackage{amsmath,graphicx,amssymb,hyperref}

\begin{document}

\title{A class of models for random hypergraphs}


\author{Marc Barthelemy}
\email{marc.barthelemy@ipht.fr}
\affiliation{Universit\'e Paris-Saclay, CEA, CNRS, Institut de Physique Th\'{e}orique, 91191, 
	Gif-sur-Yvette, France}
\affiliation{Centre d'Analyse et de Math\'ematique Sociales (CNRS/EHESS) 54 Avenue de Raspail, 75006 Paris, France}

\begin{abstract}

  Despite the recently exhibited importance of higher-order
  interactions for various processes, few flexible (null) models are
  available. In particular, most studies on hypergraphs
  focus on a small set of theoretical models. Here, we introduce a class of
  models for random hypergraphs which displays a similar level of flexibility of
complex network models and where the main ingredient is the
  probability that a node belongs to a hyperedge. When this
  probability is a constant, we obtain a random hypergraph in the same
  spirit as the Erdos-Renyi graph. This framework also allows us to
  introduce different ingredients such as the preferential attachment
  for hypergraphs, or spatial random hypergraphs. In particular, we
  show that for the Erdos-Renyi case there is a transition threshold
  scaling as $1/\sqrt{EN}$ where $N$ is the number of nodes and $E$
  the number of hyperedges. We also discuss a random geometric
  hypergraph which displays a percolation transition for a threshold
  distance scaling as $r_c^*\sim 1/\sqrt{E}$. For these various
  models, we provide results for the most interesting measures, and
  also introduce new ones in the spatial case for characterizing
  the geometrical properties of hyperedges. These different models
  might serve as benchmarks useful for analyzing empirical data.

   
\end{abstract}


\maketitle

\section{Introduction}

Complex networks became increasingly important for describing a large
number of processes and were the subject of many studies for more than
20 years now \cite{Barrat:2008,Latora:2017,Menczer:2020}. Recent
analysis of complex systems
\cite{Ghoshal2009,Battiston2020,Battiston2021,Bianconi2021} however showed that
networks provide a limited view. Indeed networks (or graphs) describe
a set of pairwise interactions and exclude any higher-order
interactions involving groups of more than two units.  With the
increasing amount of data many higher-order interactions were observed
in a large variety of context, from systems biology
\cite{Krieger2021}, face-to-face systems \cite{Cencetti:2020},
collaboration teams and networks \cite{Patania:2017,Juul:2022},
ecosystems \cite{Grilli:2017}, the human brain
\cite{Petri:2014,Giusti:2016}, document clusters in information
networks, or multicast groups in communication networks,
etc. \cite{Gao2012,Chodrow2019}. Modeling these higher-order
interactions with graphs might lead to erroneous interpretations,
calling for the need of a more flexible framework. In addition,
these higher-order interactions are highly relevant for all possible
processes that take place in these systems \cite{Battiston2021}. These
processes include contagion where a disease can spread in a non-dyadic
way \cite{Iacopini:2019,Arruda:2020,Barrat:2022}, diffusion
\cite{Schaub:2020,Carletti:2020}, cooperative processes
\cite{Alvarez:2020}, or synchronization
\cite{Skardal:2019,Millan:2020,Lucas:2020}.  Models for hypergraphs -
null, or spatial for example, are then much needed for analyzing
processes that involve the interaction between more than two nodes.

\begin{figure}[ht!]
  \includegraphics[width=0.3\textwidth]{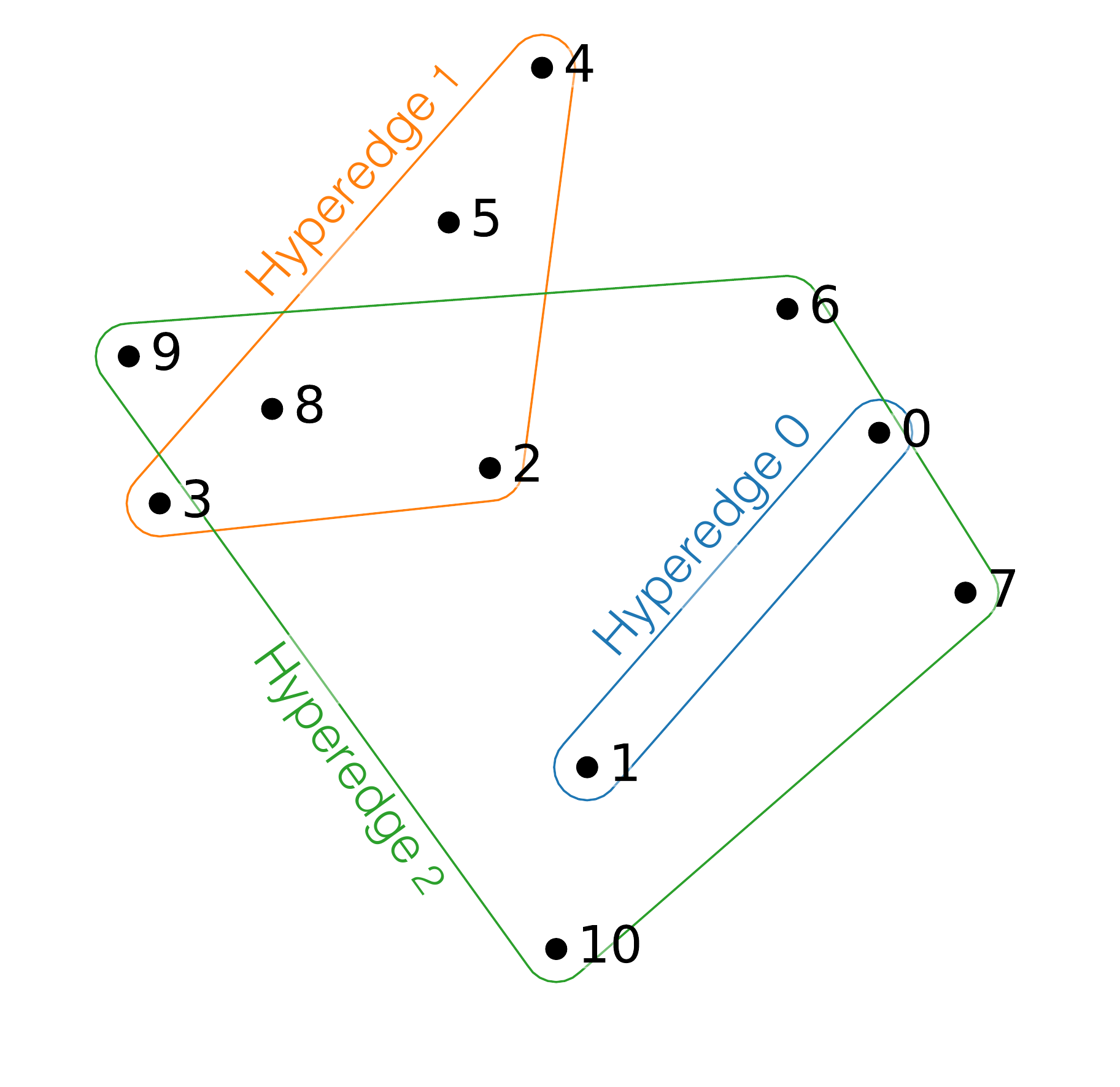}
	\caption{Classical representation of a hypergraph. We have
          here $|V|=N=10$ vertices and $E=3$ hyperedges. The
          hyperedges are $e_0=\{0,1\}$, $e_1=\{2,3,4,5,8\}$, and
          $e_2=\{0,6,7,9,10\}$, with sizes $2$, $5$, and $5$,
          respectively. All the nodes have degree $k=1$, except for
          nodes $0,1,2,8$ which have a degree $k=2$.}
        \label{fig:hypergraph_example}
 \end{figure}

In order to go beyond usual graphs, a natural
extension consists in allowing edges that can connect an
arbitrary number of nodes. These `hyperedges' constructed over a set of
vertices define what is called a hypergraph
\cite{Berge:1984,Bretto:2013}. More formally, a hypergraph $H=(V,E)$ where $V$ is a set of elements
(the vertices or nodes) and $E$ is the set of hyperedges where each
hyperedge is defined as a non-empty subset of $V$ (a simple example is shown in
Fig.~\ref{fig:hypergraph_example}).
For a directed hypergraph, the hyperedges are not sets, but
an ordered pair of subsets of $V$, constituting the
tail and head of the hyperedge \cite{Gallo:1993,Galeana:2009,Ausiello:2017}.

The number $N=|V|$ of vertices is called the order of the hypergraph,
and the number of hyperedges $M=|E|$ is usually called the size of the
hypergraph. The size of an hyperedge $|e_i|$ is the number of its
vertices. The degree of a vertex is then simply given by the number of hyperedges to which it
is connected. A simpler hypergraph considered in many studies is obtained
when all hyperedges have the same cardinality $d$ and is then called a
$d$-uniform hypergraph (the rank of a hypergraph is $r=\max_E|e|$ and
the anti-rank $\overline{r}=\min_E|e|$, and when both quantities are
equal the hypergraph is uniform). A $2$-uniform hypergraph is then a standard
graph. 

Many measures are available for hypergraphs. Walks, paths and
centrality measures can be defined and other measures such as the
clustering coefficient can be extended to hypergraphs
\cite{Zlatic:2009, Chodrow2019,Battiston2020,Joslyn2020,Aksoy2020}. A
recent study investigated the occurence of higher-order motifs
\cite{Lotito2022} and community detection was also considered
\cite{Turnbull2019,Chodrow:2021}. New measures can however be defined
and we will discuss in particular the statistics of the intersection
between two hyperedges as being the number of nodes they have in
common \cite{Chodrow:2021}. For spatial hypergraphs, the geometrical
structure of hyperedges is naturally of interest and we will discuss
some quantities that characterize it.

It was already been noted in \cite{Chodrow2019} that few flexible null
models were proposed in the context of interactions occuring within
groups of vertices of arbitrary size. In
\cite{Courtney:2017,Chodrow2019}, a null model for hypergraphs at
fixed node degree and edge dimensions is proposed. This generalizes to hypergraphs the usual
configuration model \cite{Molloy1998}. Other models for higher-order
interactions were described in the review \cite{Battiston2020} such as
bipartite models, exponential graph models, or motifs models, but in
general hypergraph models are less developed. Most of these models are
introduced in the mathematical literature and are usually thought as
immediate generalizations of classical graph models. In many of these
models, it is usually assumed that all hyperedges have the same size,
which is a strong constraint. We note that an interesting hypernetwork
growth model was proposed in \cite{Wang:2010} where both the idea of
hyperedge growth and hyperedge preferential attachment were
introduced.

Here, the goal is to present a mechanism that can generate (in a
simple way and with low complexity) a family of different hypergraphs
and which displays the same level of flexibility of complex network
models. We propose here such a framework and introduce a class of
models that relies on a single ingredient: the probability that a node
belongs to a hyperedge. We will consider various illustrations of this
class of models. We will start with the simplest one that could be
seen as some sort of `Erdos-Renyi hypergraph', and also discuss
preferential attachment.  We then introduce space in different ways
and in particular discuss in more detail a random geometric hypergraph
where the probability that a vertex belongs to a hyperedge is $1$ if
its distance is less than a threshold, $0$ otherwise. For all these
illustrations, we analyse simple measures (such as the degree of
vertices or sizes of the hyperedges), the structural transition for
the giant component, and also introduce new measures. In particular,
for spatial hypergraphs, we characterize the spatial properties of
hyperedges.

\section{A class of hypergraph models}

\begin{figure}[ht!]
  \includegraphics[width=0.4\textwidth]{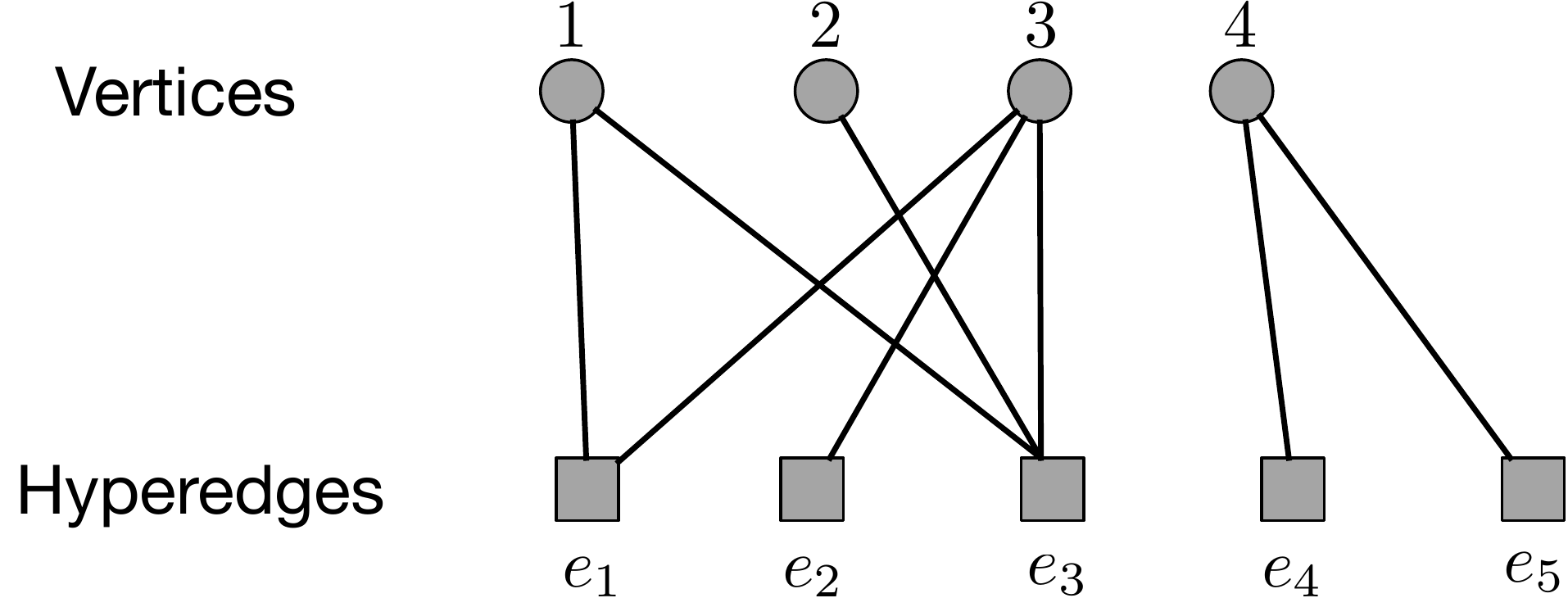}
	\caption{Example of a small hypergraph with $N=4$ vertices and $E=5$
          hyperedges represented as a bipartite graph. The degree of the vertices are $k_1=2$, $k_2=1$,
          $k_3=3$, $k_4=2$ and the sizes of the hyperedges are
          $|e_1|=2$, $|e_2|=1$, $|e_3|=3$, $|e_4|=|e_5|=1$. There are
          two connected components in this hypergraph $\{1,2,3\}$ and
          $\{4\}$. The intersection between some of the hyperedges is:
          $e_1\cap e_2=\{3\}$, $e_1\cap e_3=\{1,3\}$, etc. The sum
          $\sum_ik_i$ is equal to the sum of hyperedge sizes
          $\sum_j|e_j|$ and equal to the number of links.}
	\label{fig:hypergraph2}
      \end{figure}

Hypergraphs can be represented as bipartite graphs between the nodes
and hyperedges (see Fig.~\ref{fig:hypergraph2} for an example). In
this representation the links between nodes and hyperedges indicate a
membership relation: there is a link between node $i$ and hyperedge
$e$ if $i\in e$.
 The main idea of the class of models discussed here is to introduce
 the connection probability $P(v\in e)$ that a
 node $v$ belongs to a hyperedge $e$. This is directly related to the
 incidence matrix $I$ of the hypergraph which is a $N\times E$ matrix
 with elements $I_{ie}$ equal to $1$ if $i\in e$ and $0$
 otherwise. This connection probability can be written
 under the form
\begin{align}
	P(v\in e)=F(e,d(v,e),...)
	\label{eq:class}
\end{align}
where $F$ is in general a function of the hyperedge $e$, its
vertices, or some distance between $v$ and $e$ (we will see various
examples below). We note here that in contrast with a remark in
 \cite{Turnbull2019} that it is `unclear how to impose properties on a hypergraph when a
 bipartite representation is used', we actually believe that this
 representation provides a simple framework for introducing various
 mechanisms.  We could even think of a generalization of
 Eq.~\ref{eq:class} in the spirit of the hidden-variable model
for networks. In these models, some characteristics are assigned to
nodes such as fitnesses \cite{Bianconi:2001,Caldarelli:2002}, or coordinates
in a latent space \cite{Krioukov:2010,Serrano:2012,Kitsak:2017}. The
connection function $F$ in Eq.~\ref{eq:class} could in principle
depend on these hidden variables. For example, if each
vertex $v$ has a fitness $\eta_v$, a hyperedge composed 
of vertices $v_{i_1}, v_{i_2},\dots,v_{i_m}$ has then a fitness which
will be a function of the fitnesses of all its vertices
$\eta(e)=G(\eta_{i_1}, \eta_{i_2},\dots,\eta_{i_m})$, and the connection function could then
be chosen as
\begin{align}
  P(v\in e)=F(\eta(e))
\end{align}
In this article, we will essentially
consider the case where the function depends on some properties of the
hyperedege $e$, including its position in space. 

This definition (Eq.~\ref{eq:class}) is obviously very general and we
will focus here on different functions. First, we will consider the
constant case $F=p\in [0,1]$ which reminds us of the Erdos-Renyi model
\cite{Erdos:1960}. We will then consider the preferential attachment
case where the function $F$ depends on the size of the hyperedge. We
then end this paper by considering cases where the nodes are in a 2d
space (we will mainly consider the $d=2$ case but the generalization
to larger dimensions is trivial) and where the function $F$ depends on
a distance (to be defined) between the vertex $v$ and the edge $e$.
For all these models, we will present the result for usual measures
but also develop new measures tailored to spatial hypergraphs.

Throughout this work, we denote by $N$ the order of the hypergraph
(which is the number of nodes) and $E$ the number of hyperedges. For
all the models considered here, we will assume that the number $E$ of
hyperedges is given. Once an initial set of hyperedges is given (in
cases studied here, the initial hyperedges comprise single nodes
chosen at random), we can iterate over all nodes and apply
Eq.~\ref{eq:class}. Once all nodes are tested, we get the final
hypergraph that we measure. The simplest measure is the degree $k_i$
of a node $i$ (which is the number of hyperedges to which it
belongs). The degree can also be computed as the sum of row elements
of the incidence matrix $I_{ie}$: $k_i=\sum_eI_{ie}$ (see for example
the review \cite{Battiston2020}). The size $m_l=|e_l|$ of the
hyperedge $l$ is the number of nodes it contains. Naturally, the
average degree and size (and the distribution of $k$ and $m$) are
important quantities, and it should be noted that they are not
independent. Indeed, if we use the links in the bipartite
representation of the hypergraph (i.e. there is a link between node
$i$ and hyperedge $e$ if $i\in e$, see Fig. \ref{fig:hypergraph2}),
their number $L$ can be expressed in two different ways as
$L=\sum_i k_i=\sum_j|e_j|$. This relation can be rewritten as
\begin{align}
  N\langle k\rangle=E\langle m\rangle
  \end{align}
  where $\langle k\rangle$ denotes the average degree and $\langle
  m\rangle$ the average size of hyperedges. The distributions $P(k)$
  and $P(m)$ are then natural objects to study. We will also focus on other quantities: the intersection
between two hyperedges (a quantity that is trivial for graphs and
equal to one), and for spatial hypergraphs the spatial extension
$s(e)$ of a hyperedge $e$. We will also discuss connectivity
properties of these hypergraphs and in particular we will focus on
abrupt structural changes characterized by the emergence of a giant
component (that will need to be defined) scaling with $N$.

\section{The simplest random hypergraph}

\subsubsection{Definition}

There is no unique definition of random hypergraphs and various
generalization of the classical Erdos-Renyi graph were proposed (see
for example \cite{Juul:2022}). In particular, mathematicians like to
think of a set of all possible hypergraphs - given some parameters
(number of nodes, etc) - and to consider a uniform distribution over
this set \cite{Schmidt:1985}. More specifically, many papers focus on
$k$-uniform hypergraph for which the size of hyperedges is constant
and equal to $k$. Given a set of $V$ vertices and a set of subsets of
these vertices we can construct the natural analogue of Erdos-Renyi
graphs \cite{Erdos:1960}: each $k$-tuple of vertices is a hyperedge
with probability $p$ \cite{Cooley2015} (more formally, the set
$H^k(n, p)$ denotes the random $k$-uniform hypergraph with vertex set
$[n] = \{1, 2, . . . , n\}$ in which each of the $\binom{n}{k}$
possible edges is present independently with probability $p$). In
other studies, every hypergraph of $E$ hyperedges on $N$ nodes has the
same probability \cite{Vega:1982} (the classical Erdos-Renyi random
graph is then recovered in the case $k=2$). This type of model was
discusssed by mathematicians in particular about the phase transition
for the giant component
\cite{Schmidt:1985,Karonski:2002,Ghoshal2009,Cooley2015,Cooley2018,Kim2022}. For
$k=2$, we recover
usual graphs and from Erdos and Renyi \cite{Erdos:1960}, we know that
there is a transition for $E=cN$ with $c=1/2$ (which corresponds to an
average degree $\langle k\rangle = 2E/M=1$). This transition is
characterized by an abrupt structural change with the emergence of a
giant component scaling. The general result for hypergraphs obtained in
\cite{Karonski:2002} is similar and states that a giant component appears for
$c=1/k(k-1)$. We refer the reader interested in some mathematical aspects
of this problem to the book \cite{Frieze:2016}.

Here, we will construct a `Erdos-Renyi hypergraph' in the context of
Eq.~\ref{eq:class}. For the usual Erdos-Renyi graph, the connection
probability between two nodes is constant and similarly for the
hypergraph we will then assume that for each vertex $v\in V$ and for each hyperedge $e$, there is a
constant probability $p$ that $v$ belongs to $e$. If this probability is constant
and equal to $p$, we can then write
\begin{align}
	P(v\in e)=p
	\label{eq:er}
\end{align}
where $p\in [0,1]$. Starting from a set of $N$ nodes, we choose a
random set of $E$ hyperedges (which are $E$ nodes chosen at random),
and add recursively nodes to these hyperedges following the rule Eq.~\ref{eq:er}.

\subsubsection{Degree and size}

This random hypergraph is a very simple model and most properties are trivial. In
particular, the degree $k$ and size $m$ distributions are easy to compute and
 are binomials
 \begin{align}
   P(k_i=k)&=\binom{E}{k}p^k(1-p)^{N-k}\\
   P(|e_i|=m)&=\binom{N}{m}p^m(1-p)^{E-m}
 \end{align}
and the average degree is then $\langle k\rangle = pE$ and the average
hyperedge size $\langle m\rangle=pN$.

 
If nodes are in space - in the disk of radius $r_0$ for example (we will see below in more details examples of spatial
networks constructed over such a set), we can compute the average
spatial extent of a hyperedge $e_i$ given by
\begin{align}
  s(e_i)=\frac{1}{m_i(m_i-1)}\sum_{v_j,v_l\in e_i}d_E(v_j,v_l)
\end{align}
where $m_i=|e_i|$ is the size of the hyperedge and where $d_E(v_j,v_l)$ is
the euclidean distance between nodes $v_j$ and $v_l$. In the Erdos-Renyi
hypergraph case, the nodes of hyperedges are distributed uniformly and
the extent is given by the average distance between randomly chosen
nodes in the disk \cite{Garcia:2005}
\begin{align}
  s=s_0=\frac{128 r_0}{45\pi}
\end{align}
which is verified in our numerical simulations (figure not shown).

\subsubsection{Hyperedge intersection}

\begin{figure}[ht!]
    \includegraphics[width=0.4\textwidth]{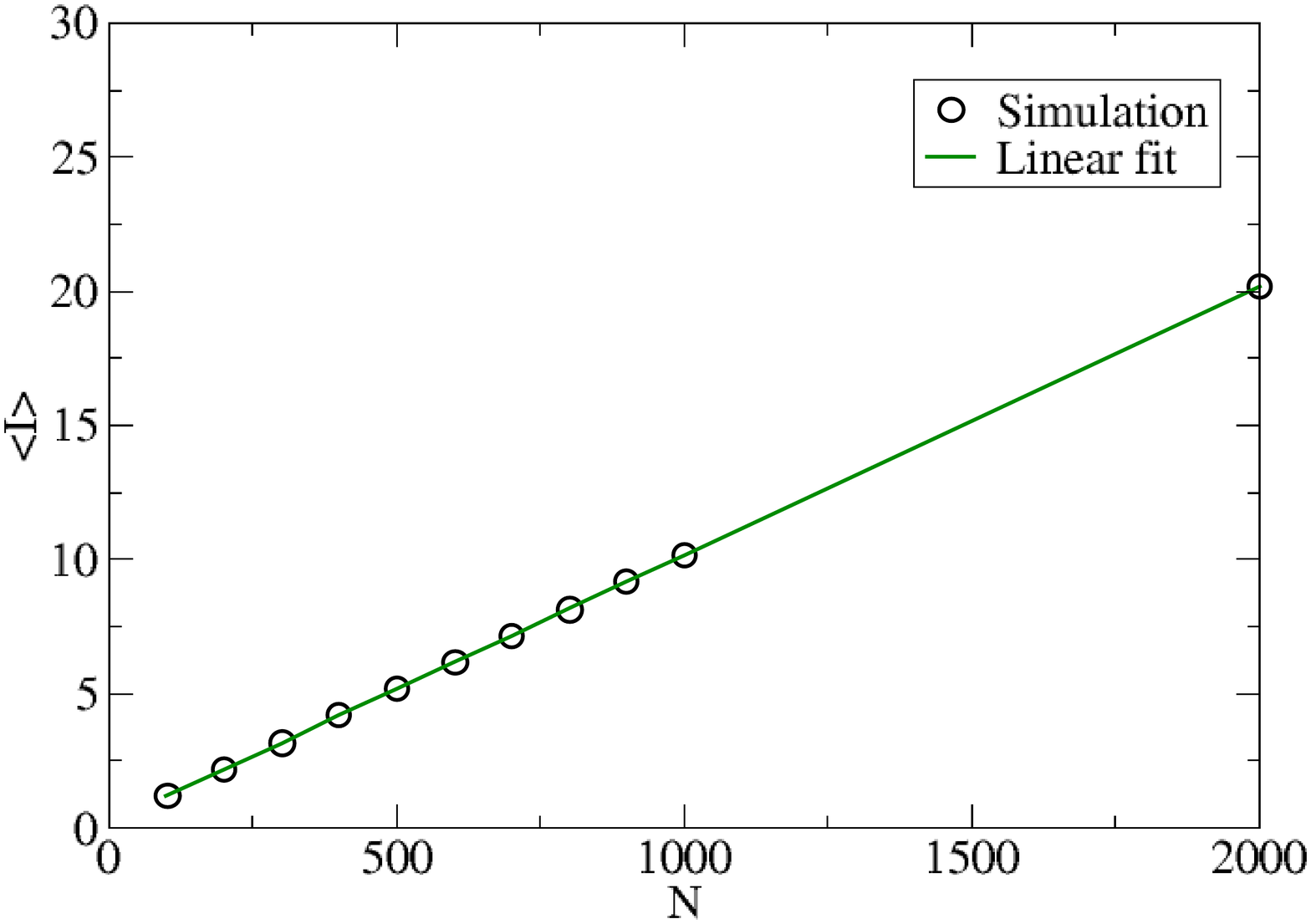}
  \includegraphics[width=0.4\textwidth]{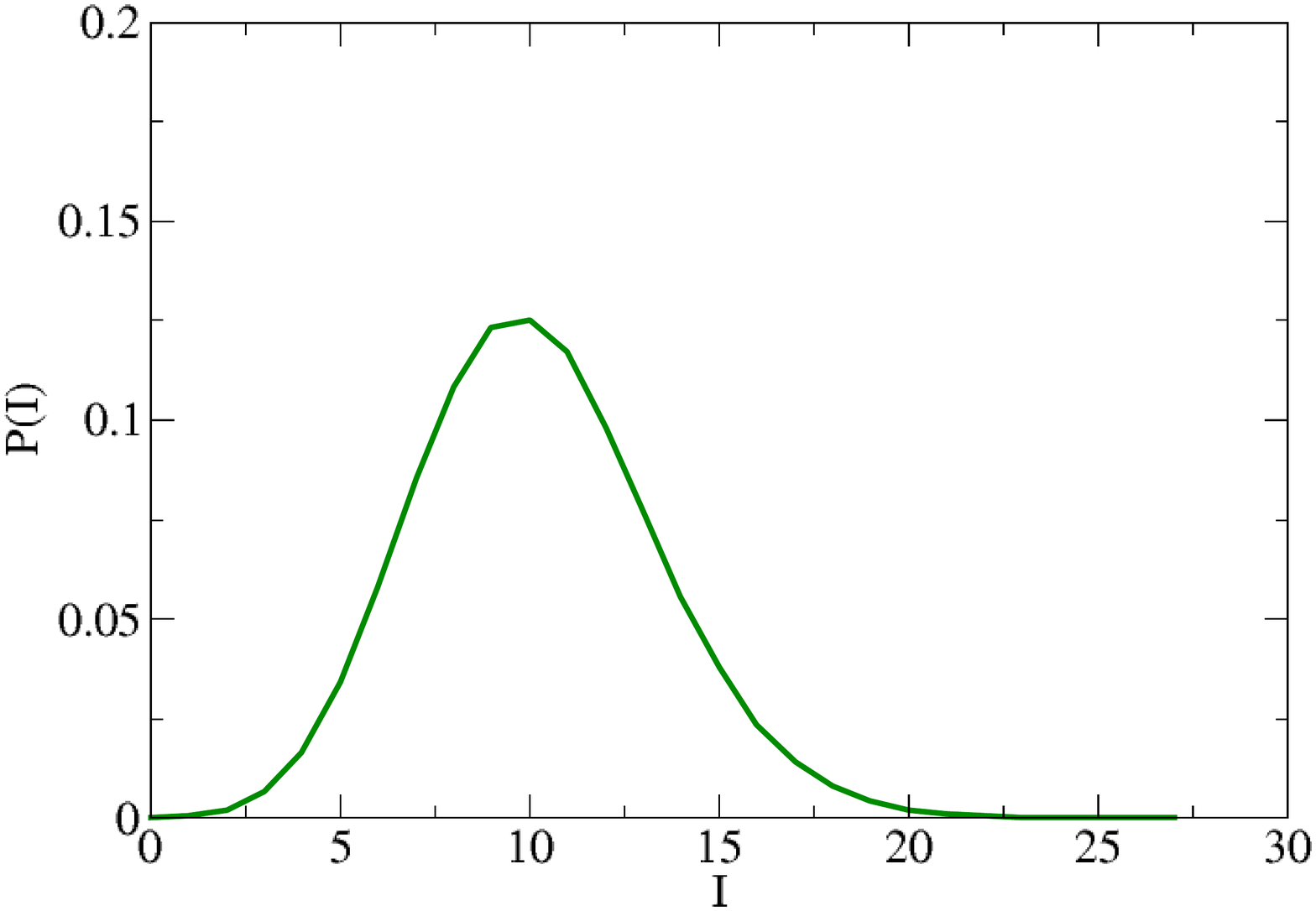}
	\caption{(a) Average
          intersection $\langle I\rangle$ versus $N$.  The line is a
          linear fit of slope $p^2$ (here $p=0.1$ and results were
          obtained by averaging over $100$ configurations). (b) Probability distribution $P(I)$ of the intersection
          between hyperedges defined in
          Eq.~\ref{eq:inter}. Simulations were obtained for $N=1000$,
          $E=100$, $p=0.1$ and averaged over $100$ configurations.}
	\label{fig:pI}
 \end{figure}

The intersection $I$
between two hyperedges $e_i$ and $e_j$ is the  number of nodes they
have in common \cite{Chodrow2020} and is denoted by
\begin{align}
  I=|e_i\cap e_j|
  \label{eq:inter}
\end{align}
where $|\cdot|$ denotes the cardinal of a set. 
For the random hypergraph discussed here, the probability that a given node belongs
 to two different hyperedges is $p^2$. The average intersection is
 then given by $\langle I\rangle=p^2N$ which we verified by
 simulations (see Fig.~\ref{fig:pI}(a)). The probability distribution of
 the intersection $I$ between two hyperedges is then a binomial of
 parameters $N$ and $p^2$
 \begin{align}
   P(I=n)=\binom{N}{n}p^{2n}(1-p^2)^{N-n}
 \end{align}
 which can be verified numerically (see Fig.~\ref{fig:pI}(b)).

\begin{figure}[ht!]
  \includegraphics[width=0.3\textwidth]{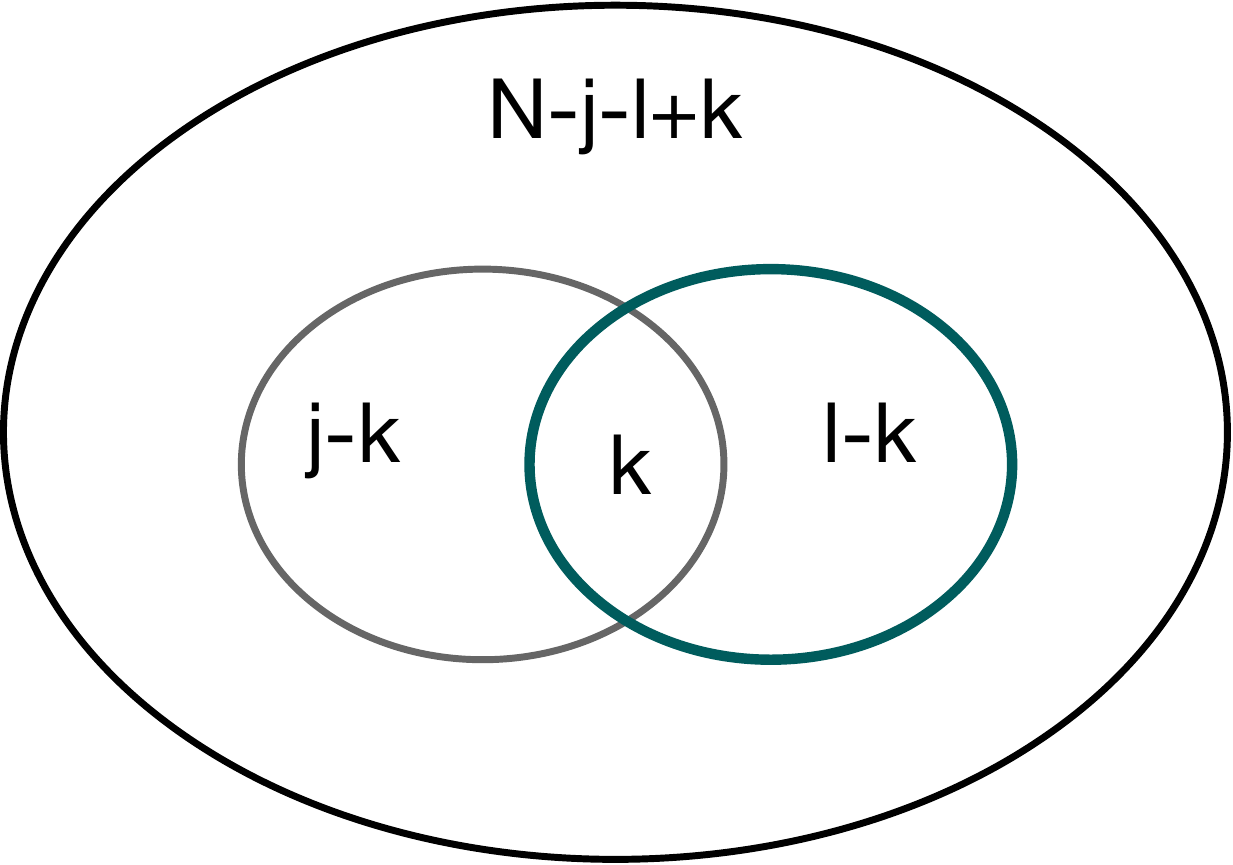}
	\caption{Schematic illustration of the intersection between
          two hyperedges of respective sizes $j$ and $l$.}
	\label{fig:inter}
\end{figure}

We can go further and define the intersection $I_{jl}$ of two hyperedges of size $j$ and
$l$, respectively \cite{Chodrow2020}. The intersection $I_{jl}$ is then a random variable
and can be expressed for the random hypergraph defined here. Indeed
the probability $P(I_{jl}=k$ is given by the number $N_{jl}(k)$ of events where
the intersection of a subset
of size $j$ and a subset of size $l$ is of size $k$ (see Fig.~\ref{fig:inter}).
The number $N_{jl}(k)$ is then given by the following multinomial coefficient
\begin{align}
  N_{jl}(k)=\frac{N!}{k!(j-k)!(l-k)!(N-j-l+k)!}
\end{align}
and the corresponding probability is
$P_{jl}(k)=N_{jl}(k)/\sum_kN_{jl}(k)$ (if needed a large $N$ analysis
could then be performed).

\subsubsection{Giant component}

We now consider the behavior of the giant component when $p$ varies. This problem actually motivated many
studies of random graphs, starting with the original paper by Erdos
and Renyi \cite{Erdos:1960}. In order to define the giant component we assume that all
nodes in the same hyperedge are connected to each other (equivalently
that each hyperedge is a complete graph) and that two hyperedges are
connected if their intersection is at least equal to one. Other
definitions are possible using the concept of high-order hypergraph walk
\cite{Aksoy2020}. A hypergraph walk is called a $s$-walk where the
order $s$ controls the minimum edge intersection over which the walk
takes place. A $1$-walk is then the usual walk and this is the
connectivity that we are using here for defining the giant component:
two nodes are connected if there is at least one $1$-walk between
them (see \cite{Aksoy2020} for a discussion about larger values of
$s$).


In the case of random hypergraphs considered here, the nodes in each
hyperedge form a connected clique and the giant component problem lies
essentially in the connection between the hyperedges. The probability
that there is at least one intersection between two hyperedges is given by
 \begin{align}
   \nonumber
   P(I\geq 1)&=1-(1-p^2)^N\\
   &\approx p^2N
 \end{align}
for $p^2N\ll 1$. If the number of hyperedges $E$ is large, there is a
giant component  in the hypergraph if the $E$ hyperedges constitute a
giant component. The classical result for Erdos-Renyi random graph
states that there is a giant component appearing at average degree
equal to $1$ which leads here to
 the condition $p_c^2NE=1$. The threshold then behaves as 
 \begin{align}
   p_c\sim\frac{1}{\sqrt{NE}}
 \end{align}
We note that this result obtained by a simple argument has been already found by
more rigorous methods for random bipartite graphs in \cite{Johansson2012}.

\section{Preferential attachment for hypergraphs}


The preferential model for networks states that the probability that a
new node $n$ connects to an existing one $i$ is proportional to the
degree $k_i$ of $i$ \cite{Barabasi:1999}. It might be interesting to
extend this reinforcement mechanism to hypergraphs. Few studies
discussed this apart from the notable exception of
\cite{Wang:2010} and \cite{Bianconi:2017,Courtney:2017,Sun:2021}. In the model proposed by
\cite{Wang:2010}, two important ingredients were introduced for the growth of a random
hypergraph. First, at every time step, a new hyperedge $e$ is constructed
(either with $m$ new nodes and a randomly chosen node in the existing
hypergraph or from a random number of nodes selected at random in the
existing hypergraph and a new node such as in \cite{Yang:2013}). Second, they introduced a hyperedge preferential
attachment where the hyperedge $e$ is connected to an existing node
$i$ with probability proportional to the degree of $i$ (also called
hyperdegree in this study) \cite{Wang:2010}. This was generalized in
\cite{Guo:2016}. In \cite{Wang:2010}, it was shown that the
hypergraph constructed in this way shares many similar features with
complex networks such as scale-free property of the degree
distribution, etc. \cite{Wang:2010,Yang:2013}.

Clearly, there are several ways to introduce preferential attachment
in the hypergraph formation, but in the framework defined by
Eq.~\ref{eq:class}, the natural choice is to write the connection probability
as a function of the hyperedge size $m=|e|$
\begin{align}
	P(v\in e)=F[|e|]
\end{align}
and the simplest function is the linear one
\begin{align}
  P(v\in e_i)=\frac{|e_i|}{\sum_j|e_j|}
  \label{eq:prefattach}
\end{align}
which introduces a rich-get-richer process through the size of
hyperedges. Other choices could include for example the average degree
of nodes contained in the hyperedge $e$, etc.

In this simple model, the degree of each vertex is $k=1$ (a simple
generalization
would consist in taking for each vertex $n$ possible connections to different
hyperedges).  The probability that a given vertex connects to a given
hyperedge is given by Eq.~\ref{eq:prefattach} and this problem is
exactly a Polya urn with
$E$ colors. The limiting distribution for large time (and therefore at
large number of nodes $N$) can be shown to be the Dirichlet
multinomial distribution \cite{Mahmoud:2008}
\begin{align}
  \nonumber
  P_t(m_1,m_2,\dots,m_E|\alpha_1,\alpha_2,\dots,\alpha_E)&=\\
  \frac{\Gamma(\alpha_0)\Gamma(N+1)}{\Gamma(N+\alpha_0)}
  \prod_{j=1}^E
  &\frac{\Gamma(m_j+\alpha_j)}{\Gamma(\alpha_j)\Gamma(m_j+1)}
\label{eq:dirich}
\end{align}
where $m_j=|e_j|$, $\alpha_j$ denotes the initial size of hyperedge $j$,
$\alpha_0=\sum\alpha_m$ and $\Gamma(x)$ is the gamma function.  In
particular, if we start with all the hyperedges having the same size,
the limiting distribution is uniform. More generally, it shows that in
this case the limiting distribution depends crucially on the initial
structure of hyperedges, which certainly renders the empirical
identification of a preferential attachment mechanism difficult.

\begin{figure}[hbt!]
  \includegraphics[width=0.5\textwidth]{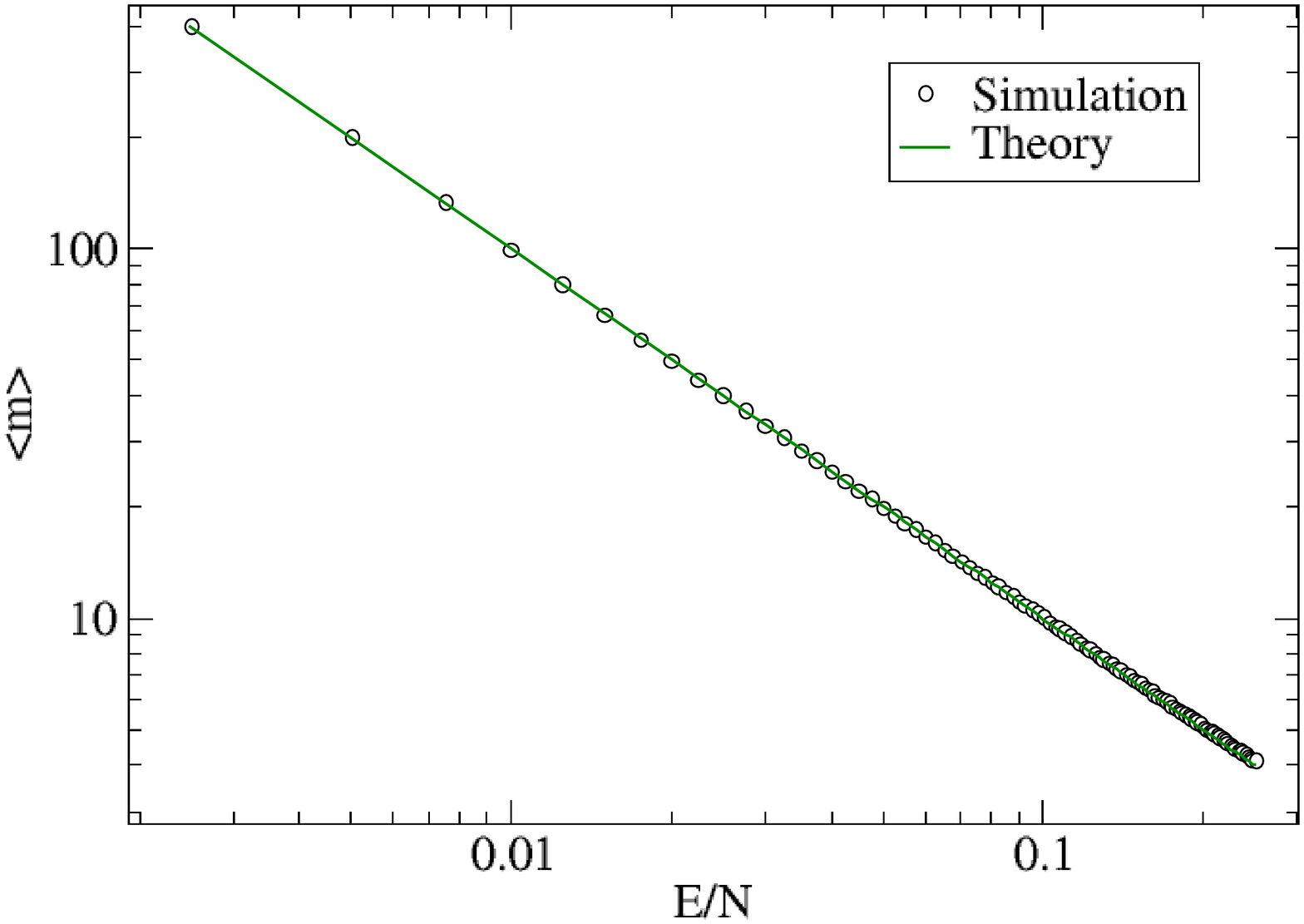}
  \includegraphics[width=0.5\textwidth]{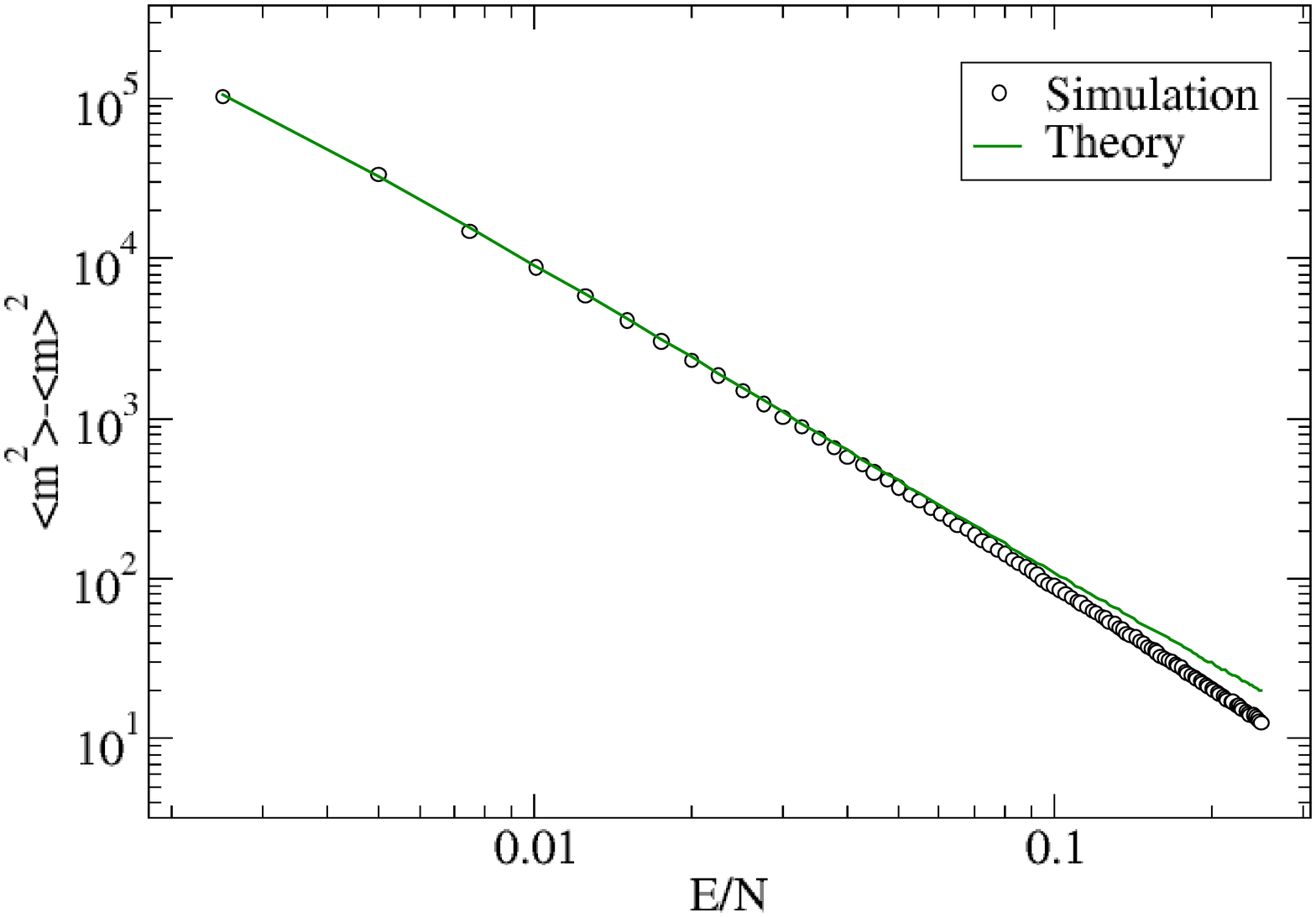}
	\caption{Comparison of simulations and theoretical results
          obtained from the multinomial Dirichlet distribution
          Eq.~\ref{eq:dirich} for the uniform case. The symbols are obtained for numerical simulations
          for $N=1000$ (averages are computed over $1000$ configurations) and the
          line represent the theoretical result Eq. \ref{eq:avem}, \ref{eq:varm}.
          (a) Average hyperedge size versus the number $E$ of
          hyperedge. (b)  Variance of the hyperedge size versus $E$.}
 \label{fig:moments}
\end{figure}

This multinomial Dirichlet distribution is a bit difficult to test but
we can easily compute its first two cumulants. Indeed for an initial
condition given by $\alpha_1,\alpha_2,\dots,\alpha_E$, the average
size of edge $m_i$ is given by
\begin{align}
  \langle m_i\rangle = N\frac{\alpha_i}{\alpha_0}
  \label{eq:avem}
\end{align}
For the variance, we note that the marginal distribution is a
Beta-Binomial which then leads to the result
\begin{align}
  \mathrm{Var}(m_i)=
  N\frac{\alpha_i}{\alpha_0}
  \left(
  1-\frac{\alpha_i}{\alpha_0}
  \right)
  \frac{N+\alpha_0}{1+\alpha_0}
  \label{eq:varm}
\end{align}
In the uniform case (all the $\alpha_i$s are equal to some value $\alpha$), we then obtain
$\langle m_i\rangle = N/E$, and
$\mathrm{Var}(m)=N/E(1-1/E)(N+E\alpha)(1+E\alpha)$. We can test this
  result by varying $E$ for example and we obtain the results shown in
  Fig.~\ref{fig:moments}. The agreement between the numerical
  simulation and these results is excellent in the case of the
  average (Fig.~\ref{fig:moments}(a)). In the case of the variance, we
  observe a small deviation from the theoretical result for $E$ becoming
  closer to $N$. In this case, the average size of the hyperedge can
  be small (of order $N/E$) and fluctuations can be large (the
  deviation then decreases with the number of configurations).


\section{Random spatial hypergraphs}

\begin{figure}[ht!]
  \includegraphics[width=0.3\textwidth]{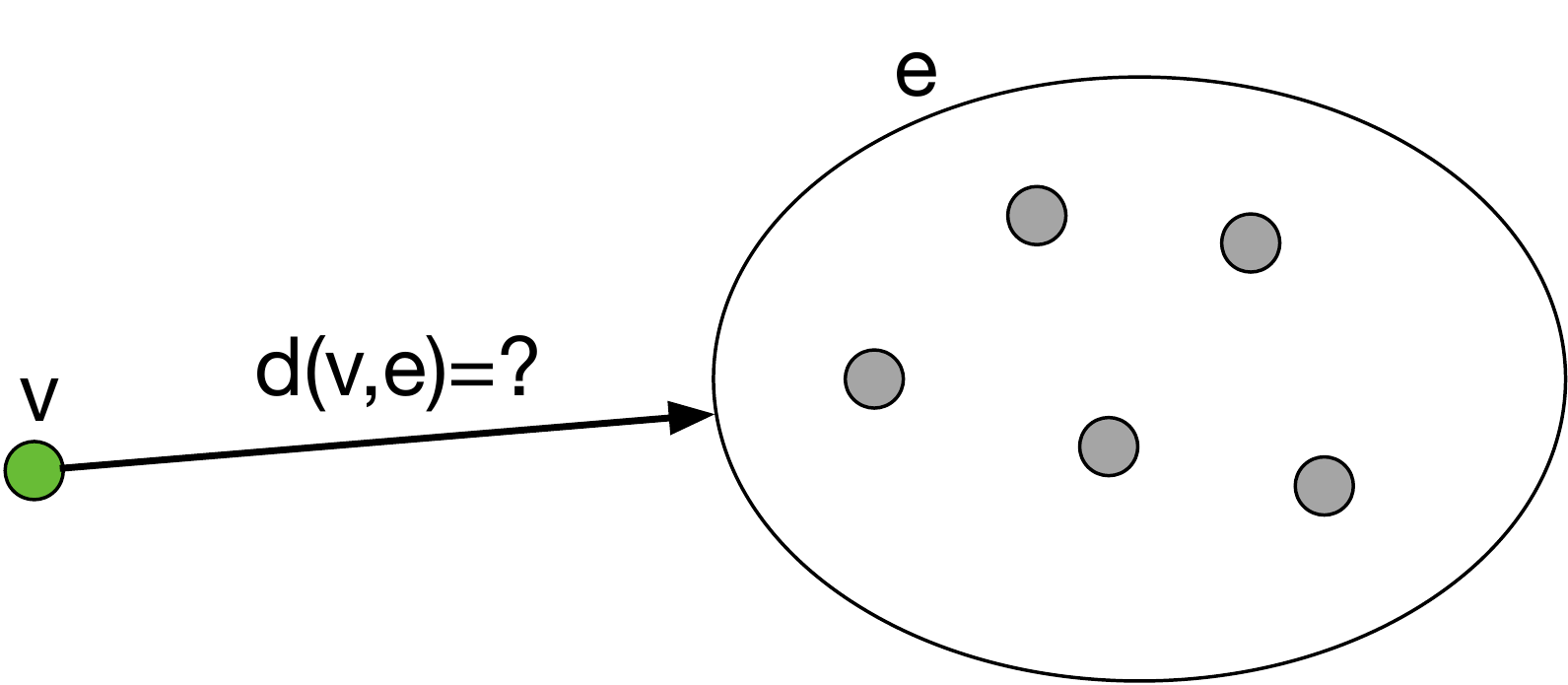}
	\caption{Schematic illustration of the choice when
          constructing a spatial hypergraph. A new node $v$ enters the
          hypergraph and will connect to the hyperedge $e$. The
          problem is how to compute the distance $d(v,e)$ between the
          node and the hyperedge.}
	\label{fig:spahg}
      \end{figure}

It is reasonable to think that in some instances, introducing space is
necessary. For example, contagion among a group of individuals
naturally involves space through the proximity needed to transmit an
infectious disease. Other systems where space is relevant
(communications networks, neural networks, etc) and where higher-order
interactions take place can be described by what we could call
`spatial hypergraphs' and we will discuss here some simple examples of such objects,

We assume that the $N$ nodes are distributed uniformly on a disk
of radius $r_0$. Each node $i$ has a position $x_i$ and it is then
natural to consider a model defined by the following connection probability 
\begin{align}
	P(v\in e)=F[d(v,e)]
\end{align}
where $F$ is a given function, and $d(v,e)$ measures the distance
between the vertex $v$ and the hyperedge $e$. There are
many choices for defining $d(v,e)$ and many models are possible
(Fig.~\ref{fig:spahg}).

It is obviously hopeless (and probably useless too) to try to explore all
possible cases, and we will focus on two main models. First, we will
assume that the function $F$ decreases with the distance as an
exponential $F(d)\sim \mathrm{exp}(-d/r_c)$. We will then consider a
model close - in spirit at least - to the random geometric graph \cite{Gilbert:1959}.

\subsection{Exponential case}


We consider the case where $F$ is an exponential function decreasing
with distance (we expect similar results for other decreasing functions). This corresponds to the intuitive idea that it is more
difficult for a node to belong to a distant hyperedge. The range of
the exponential is denoted by $r_c$ and the connection probability
then reads as
\begin{align}
	P(v\in e)=p\mathrm{e}^{-d(v,e)/r_c}
\end{align}
where $p\in [0,1]$ and $d(v,e)$ is a measure of the distance between
the node $v$ and the hyperedge $e$. When $r_c\gg r_0$, the exponential term is
essentially $1$, space is then irrelevant and we recover the random
hypergraph model discussed above. For the distance $d(v,e)$, many choices are possible,
such as the average over all nodes, the minimum or maximum distance
among the nodes, and we will essentially 
present the results for the average distance
\begin{align}
  d(v,e)=\frac{1}{m}\sum_{w\in e}d_E(v,w)
  \label{eq:dave}
\end{align}
where $m=|e|$ is the size of the hyperedge $e$ and $d_E(v,w)$ is the
euclidean distance between $v$ and $w$.

\begin{figure}[ht!]
  \includegraphics[width=0.4\textwidth]{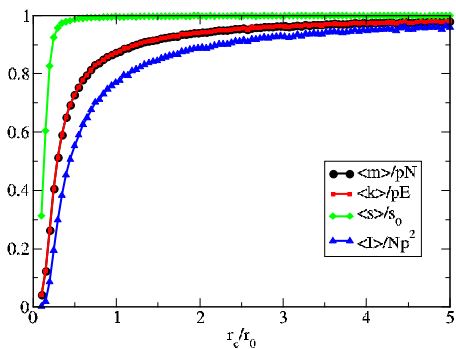}
	\caption{Average degree $\langle k\rangle$, hyperedge size $m$, intersection $\langle
          I\rangle$, and hyperedge extent $\langle s\rangle$ versus
          $r_c/r_0$ and normalized by their value for the random
          hypergraph (for
          $p=0.1$, $N\in [100,1000]$, $E\in [100,N]$, $100$ configurations).}
	\label{fig:all_vs_rc}
\end{figure}

For $r_c\gg r_0$ we recover the random hypergraph for which we know
most quantities. We can then rescale the average degree by $pE$, the
average hyperedge size by $pN$, the average intersection by $Np^2$ and
the average spatial extent of hyperedges by $128r_0/45\pi$. These
quantities versus $r_c$ are shown in Fig.~\ref{fig:all_vs_rc}
The total number of links is given by $\sum_jk_j=\sum_i|e_i|$ which
implies that $\langle m\rangle/pN=\langle k\rangle/pE$ as observed in
Fig.~\ref{fig:all_vs_rc}. Also and as expected, all these quantities grow with $r_c$, but at different
speeds.

We can consider other choices for the distance $d(v,e)$ instead of
Eq.~\ref{eq:dave}. For example, the minimum distance is also a
reasonable choice that can make sense for some systems
\begin{align}
  d(v,e)=\min_{w\in e}d_E(v,w)
  \label{eq:dmin}
\end{align}
Also, in the case where we define the centroid $c(e)$ of the hyperedge with
 coordinates $(x_c(e),y_c(e))=1/|e|\sum_{i\in e}(x(i),y(i))$, the
 distance can be computed from this centroid
 \begin{align}
   d(v,e)=d(v,c(e))
   \label{eq:dg}
 \end{align}
The corresponding hypergraph model bears some similarity with
 the k-means clustering method (see for example \cite{Jain:1999}). This method partitions $N$ observations into
$k$ clusters, with the rule that each node belongs to the cluster with
the nearest centroid location. In the hypergraph case, each time a new node is attached
to a hyperedge, the centroid position is recalculated, as in the Lloyd
algorithm \cite{Lloyd:1982}. We averaged here over all possible initial positions of the
hyperedges, but as in the k-means clustering case, probably some
further work is needed in order to understand the effect of initialization on the
resulting hypergraph structure.

\begin{figure}[ht!]
  \includegraphics[width=0.4\textwidth]{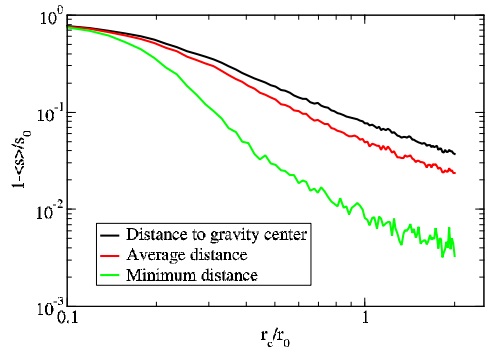}
	\caption{One minus the average normalized spatial extent of
          hyperedges $1-\langle s\rangle/s_0$  versus $r_c$ shown in
          loglog (results are obtained for $E=100$, $N=1000$ and $100$
          configurations). A power law fit over the last decade gives
          an exponent going from $1.2$ to $1.5$ for the different cases.}
	\label{fig:ps_all}
      \end{figure}

We compare the convergence to the random case of the average spatial extent
of hyperedges $\langle s\rangle$ for these 3 different choices Eq.~\ref{eq:dave},
\ref{eq:dmin}, \ref{eq:dg} and the results for $\langle s\rangle$ versus $r_c$ are shown in Fig.~\ref{fig:ps_all}.
We observe different speeds to convergence to the random case. In
particular, the cases Eq.~\ref{eq:dave} and \ref{eq:dg} behave
similarly, while the case Eq.~\ref{eq:dmin} seems to converge in a
much faster way. The average spatial extent of hyperedges thus seems
to be very sensitive to the choice of the connection probability.


 
 \subsection{A random geometric hypergraph}

 \subsubsection{Definition}
 
 The random geometric graph is a classical spatial graph introduced by
 Gilbert \cite{Gilbert:1959} where nodes
 are located in the 2d plane and are connected if their distance is
 less than a threshold $r_c$ (for mathematical properties of this
 object, see \cite{Penrose:2003}). If we denote by $\rho=N/A$ the density
 of nodes in the area $A$, the average degree is given by
 \begin{align}
   \langle k\rangle = \rho\pi r_c^2
 \end{align}
 There is a critical value $k_c$  for this quantity above which
 there is a giant cluster of size of order $N$. The value of $k_c$ is
 not exactly known but is approximately given by $k_c\approx 4.5$ (see
 for example \cite{Barthelemy:2022} and references therein). 

 There are various possibilities to extend to hypergraphs this idea of
 the random geometric graph. For example, in \cite{Turnbull2019}, the authors added hyperedges connecting all nodes that are at a distance less
 than a threshold $r_c$. Varying this threshold $r_c$ gives then a
 sequence of hyperedges included in each other. 
%
%
%
%
%
 In the framework discussed here, we choose for the connection function the following form
\begin{align}
  P(v\in e)=\theta(r_c-d(v,e))
  \label{eq:def_rgg}
\end{align}
where $d(v,e)$ is a distance between the vertex $v$ and the hyperedge
$e$ ($\theta(x)$ is the Heaviside function). As in the previous
section, there are several possible choices for defining $d(v,e)$ and we choose here
\begin{align}
  d(v,e)=\max_{w\in e}d(v,w)
\end{align}
This choice corresponds to the intuitive idea that
a vertex belongs to an hyperedge if all of its vertices are close
enough (and at a distance less than $r_c$). 

\subsubsection{Average degree and size}

\begin{figure}[ht!]
  \includegraphics[width=0.45\textwidth]{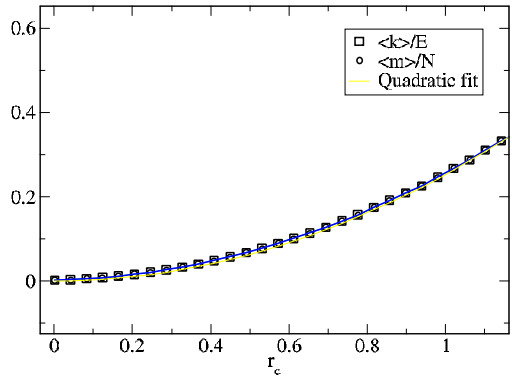}
	\caption{Average hyperedge size $\langle m\rangle/N$ and degree $\langle
          k\rangle/E$ normalized by the
          number of nodes  and the number of hyperedges,
          respectively. The line is a quadratic fit of the form $ar_c^2$
          where $a=0.25$ here ($N=1000$, $E=100$, $100$ configurations).}
	\label{fig:rgg_kave_mave}
\end{figure}

Some properties of this `random geometric hypergraph' can be discussed
with simple scaling arguments. First, the probability that a vertex belongs to an hyperedge is
(for a uniform distribution of nodes) given by $\pi r_c^2/\pi r_0^2$ (here the size
of the disk is $r_0=1$). The average degree is then
\begin{align}
  \langle k\rangle =Er_c^2
\end{align}
Similarly, the average size of the hyperedges is given by the number of nodes in
their vicinity at a distance less than $r_c$. Their size is then
simply given by
\begin{align}
  \langle m\rangle\approx \rho r_c^2\sim Nr_c^2
\end{align}
These results are consistent with the general relation $\langle
m\rangle/N=\langle k\rangle/E$, and imply a behavior $r_c^2$. Both
these results are perfectly verified in numerical simulations (in
Fig. \ref{fig:rgg_kave_mave} we show the quadratic fit).

\subsubsection{Hyperedge intersection}

\begin{figure}[ht!]
  \includegraphics[width=0.45\textwidth]{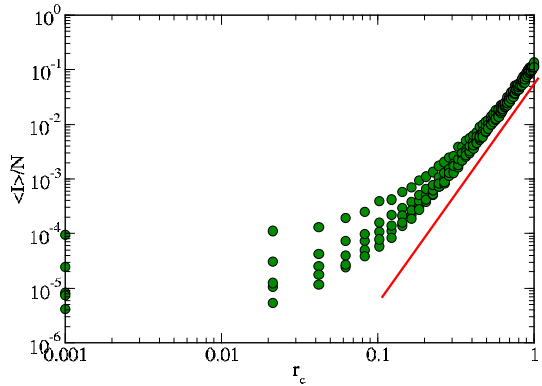}
	\caption{Average intersection $\langle I\rangle$ rescaled by
          $N$ versus $r_c$ for various values of $N$ from $200$ to
          $1000$ (here $E=100$). We observe for large $r_c\gg
          1/\sqrt{E}$, a good collapse behaving as $r_c^2$. The
          straight line meant as a guide to the eye is a power law with exponent $2$.}
	\label{fig:Iave_vs_rc}
\end{figure}

The average extent of a hyperedge is here trivially given by $\langle s\rangle
\approx r_c$ and is not a very interesting measure here. More
interesting is the intersection $I$ between two hyperedges. In order
to estimate this quantity, we first consider the area
$A(r,\ell)$ defined
by the intersection of two disks with the same radius $r$ and
separated by a distance $\ell$. Its expression can be found by
elementary geometry
\begin{align}
  A(r,\ell)=2\cos^{-1}(\ell/2r) r^2-\frac{\ell}{2}\sqrt{4r^2-\ell^2}
\end{align}
for $\ell\leq 2r$ and $A=0$ for $\ell>2r$. The probability $p_I$ that a node
belongs to the intersection of two hyperedges separated by a typical
distance $\ell\sim 1/\sqrt{E}$ is then given by
\begin{align}
  p_I\approx \frac{A(r_c,1/\sqrt{E})}{\pi r_0^2}
\end{align}
and the average intersection is $\langle I\rangle=p_I N$. It is
 $0$ for $r_c \lesssim 1/\sqrt{E}$ and for large $r_c$ behaves as $\langle
 I\rangle\sim Nr_c^2$. We thus plot $\langle I\rangle/N$ versus $r_c$
 and we indeed observe a quadratic behavior for large $r_c$ (see Fig.~\ref{fig:Iave_vs_rc}). 



\subsubsection{Giant component: transition}

\begin{figure}[ht!]
   \includegraphics[width=0.45\textwidth]{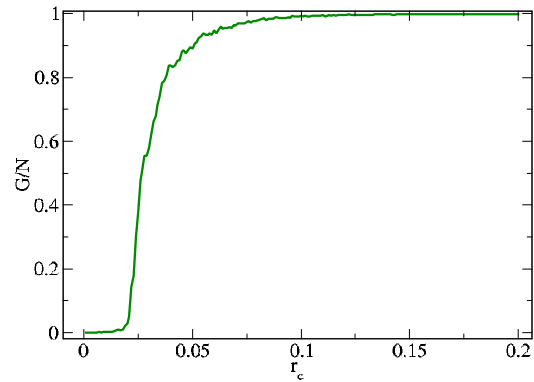}
   \includegraphics[width=0.45\textwidth]{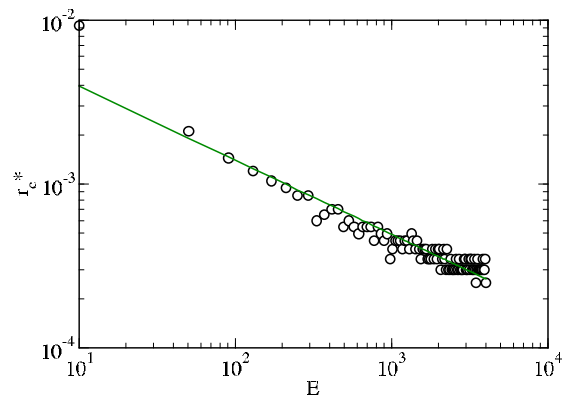}
	\caption{(a) Size $G$ of the giant component normalized by the
          number of nodes $N$ versus the interaction range $r_c$ for
          the random geometric hypergraph defined in
          Eq. \ref{eq:def_rgg} (here $N=2000$, $E=100$, and averaged
          over $100$ configurations). (b) Critical radius value $r_c^*$ versus $E$. The line is
          a power law fit of the form $r_c^*\sim 1/E^\eta$ with
          $\eta\approx 0.45$ ($r^2=0.95$). Simulations are done for
          $100$ configurations and $N=5000$.}
	\label{fig:gc_rgg}
 \end{figure}

As discussed for the random hypergraph defined above, we need a
definition for the connectivity in order to compute the giant
component. As above, we will consider $1$-walks and that all
nodes in the same hyperedge are connected to each other (equivalently
that each hyperedge is a clique or a complete subgraph) and that two hyperedges are
connected if their intersection is at least equal to one. With this
definition, we can compute the largest component and see how it varies
with $r_c$. We obtain the result shown in Fig.~\ref{fig:gc_rgg} (a)
displaying an abrupt transition for a value $r_c=r_c^*$. In order to
estimate this threshold $r_c^*$, we propose the following argument.
The hyperedges can be seen as different clusters of size $r_c$, and
the existence of a giant
component can then be mapped to the problem of continuum percolation
of $E$ disks of radius $r_c$. It is well known (see for example \cite{Meester:1996})
that percolation in this case is reached for
\begin{align}
  \rho_D a = \eta_c
\end{align}
where $\rho_D=E/A$ is the density of disk (here $A$ is the total area
given by $A=\pi r_0^2$) and $a=\pi r_c^2$ is the area of the
disks. The threshold quantity $\eta_c$ has been estimated numerically
and is approximately $\eta_c\approx 1.12$ for 2d continuum percolation
(see for example
\cite{Mertens:2012}). The critical value for $r_c^*$ is then behaving for large $E$ as
\begin{align}
  r_c^*\sim \frac{r_0}{\sqrt{E}}
\end{align}
This result is consistent with simulations shown in Fig.~\ref{fig:gc_rgg}(b).

\section{Discussion}

The observed relevance of higher-order interactions in empirical data
pushed the scientists interested in complex systems to go beyond usual
graphs and to consider random hypergraphs (or other models). It could even be possible
that in the future, we have to extend hypergraphs to multilayered
structures as discussed in \cite{Sun:2021,Vazquez:2022}. The literature about
hypergraph modeling is very heterogeneous, and sometimes difficult
to grasp, and didn't reach the state-of-the-art observed for complex
networks. 

Here, we contributed to the modeling of these higher-order
interactions and explored a particular class of random hypergraphs
where the number of hyperedges is given and where their size is
determined by some sort of hidden-variable modeling. Many alternatives
are certainly possible, but the main avantadge of this framework is
its flexibility (with the drawback of fixing the number of hyperedges,
a constraint that could probably be lift off in future models). An
important purpose of this article is to highlight the vast space of
possible hypergraph models that are left to be explored.

Many directions for future studies can be envisioned. In particular,
for spatial hypergraphs, it would be interesting to generalize the
standard models of graphs such as the Gabriel or Delaunay graphs,
beta-skeletons, etc.  Such models could in particular be helpful for
understanding the impact of space on some processes over hypergraphs
such as contagion or diffusion for example. It would also be
interesting to consider hypergraphs models based on optimal
considerations. For example, can we construct the equivalent of the
minimum spanning tree for spatial hypergraphs, or more generally, can
we define optimal hypergraphs? The field of hypergraphs is certainly
not as mature as complex networks but the recently revealed interest
in these higher-order interaction structures will cetainly trigger
many interesting studies and we can hope to see beautiful results in
the future.


\section*{Acknowledgments}
I warmly thank Ginestra Bianconi, Jean-Marc Luck, Kirone Mallick, and Erwan Taillanter for stimulating
discussions.

\bibliographystyle{prsty}

\end{document}